# Hardware Aspects, Modularity and Integration of an Event Mode Data Acquisition and Instrument Control for the European Spallation Source (ESS)

**T Gahl**[1,5]**, M Hagen**[1]**, R Hall-Wilton**[1,2]**, S Kolya**[1]**, M Koennecke**[3]**, M Rescic**[1]**, T H Rod**[1]**, I Sutton**[1]**, G Trahern**[1] **and O Kirstein**[1,4]

[1] European Spallation Source, PO Box 176, SE-22100 Lund, Sweden
[2] Mid-Sweden University, SE-85170 Sundsvall, Sweden
[3] Laboratory for Neutron Scattering and Imaging, Paul Scherrer Institut, CH-5232 Villigen PSI, Switzerland
[4] University of Newcastle, Callaghan, Australia

E-mail: thomas.gahl@esss.se

**Abstract.** The European Spallation Source (ESS) in Lund, Sweden is just entering the construction phase with 3 neutron instruments having started in its design concept phase in 2014. As a collaboration of 17 European countries the majority of hardware devices for neutron instrumentation will be provided in-kind. This presents numerous technical and organisational challenges for the construction and the integration of the instruments into the facility wide infrastructure; notably the EPICS control network with standardised hardware interfaces and the facilities absolute timing system. Additionally the new generation of pulsed source requires a new complexity and flexibility of instrumentation to fully exploit its opportunities. In this contribution we present a strategy for the modularity of the instrument hardware with well-defined standardized functionality and control & data interfaces integrating into EPICS and the facilities timing system. It allows for in-kind contribution of dedicated modules for each instrument (horizontal approach) as well as of whole instruments (vertical approach). Key point of the strategy is the time stamping of all readings from the instruments control electronics extending the event mode data acquisition from neutron events to all metadata. This gives the control software the flexibility necessary to adapt the functionality of the instruments to the demands of each single experiment. We present the advantages of that approach for operation and diagnostics and discuss additional hardware requirements necessary.

## 1. Introduction
The European Spallation Source (ESS) is designed as a long pulse neutron source with the maximum overall flux as its main objective. Integrated flux levels will be much higher than existing facilities and the geographical layout will comprise instruments of 160m and longer spanning over 3 instruments halls. At the same time the instruments for these kinds of sources need to be more flexible and complex with up to 20 choppers along the beam line. All this presents operational challenges that are best addressed with a good mixture of techniques from existing neutron facilities and other disciplines like x-ray experiments, fundamental physics or industrial applications. Advanced grounding concepts, high rate data acquisition, flexible experimental setup, advanced remote diagnostics tools, pre-emptive maintenance are a few keywords for that.

---

[5] To whom any correspondence should be addressed.

At the same time the ESS project represents also an organisational challenge with most of the instruments hardware provided in-kind by the 17 European partner countries. Integration of these contributions either as single instruments components or complete instruments into the ESS infrastructure is only possible with a flexible and modular instrument control system. This also supports clear definitions of functionalities and interfaces of the modules. We will present our ideas of such a system addressing technical, operational and organisational challenges of the future ESS.

## 2. ESS control infrastructure

### 2.1. EPICS control system (ICS)
ESS will use EPICS as a global control system for instruments, machine (accelerator, target), central infrastructure and part of conventional facilities. It acts as a horizontal device layer for data exchange between the different parts of the facility although for instrument control and operation the main signal flow will be vertically between the instruments hardware below and the user interfaces on top of the layer (figure 1). All ESS hardware links to the EPICS layer by means of standardised input/output systems (control boxes). The Integrated Controls System Division (ICS) will provide control boxes, EPICS network and hardware driver all over the facility. This ensures a high degree of standardisation and synergy within ESS.

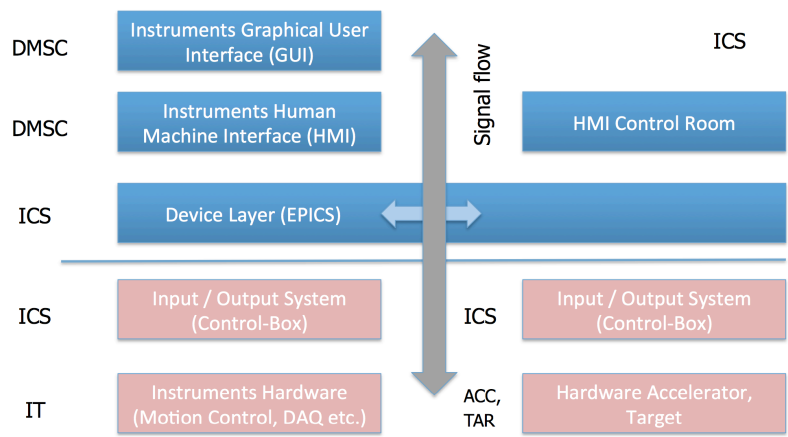

**Figure 1.** ESS facility wide control system layers (blue: software, rose: hardware).

### 2.2. User interface (DMSC)
All user interface scripts/GUIs as part of the software layers on top of EPICS are provided by the Data Management and Software Centre (DMSC). The entire control loop of the experiment including scientific calculations, coordinate transformations, sequencing, local experiment monitoring, and all user interface is handled within this layers. This includes data acquisition and data reduction for monitoring purposes as well. The DMSC software interfaces to EPICS and for large data volumes directly with the instruments hardware.

It should be noted that nothing is configured locally 'on the instrument'. This makes the DMSC interface the only access for the instruments users; there is nothing like a local monitoring at the instrument. The only local systems are for expert diagnostics and commissioning. The instrument cannot perform normally (even in a limited fashion) without DMSC supervision.

### 2.3. Timing network and signals (ICS)
ESS is using a centralised absolute timing system that is provided by ICS [1]. A generator is distributing the clock of a master oscillator and the absolute facility time to a number of timing receivers through a dedicated optical fibre network. A transport layer solution presented by Micro Research Finland [2] is envisaged as technical solution. Per default a timing receiver is present in every control box and thus makes the functionality available to all hardware connected to EPICS. Delay times in the network are compensated in steps of 11.357ns (equals 2.4m fibre length) once the exact position of the receiver and thus the length of the fibre are fixed.

Timestamps are global, across the entire facility, although the resolution of the time stamp used can be optimised in each instance; the highest resolution is 11.357ns. Coordinated synchronous strobes are available at the timing receiver interface (figure 2) to facilitate coordinated synchronous control across instrument subsystems. So, for example, instrument run control can be coordinated to any required precision. Signals include 88.0525Mhz master clock, custom clocks in fractions of the master, 14Hz master pulse, and any custom pulses in multiples and fractions of 14 Hz; all synchronised with jitter requirements as low as 1ns and possibly delayed to the master pulse in multiples of 113,57ns.

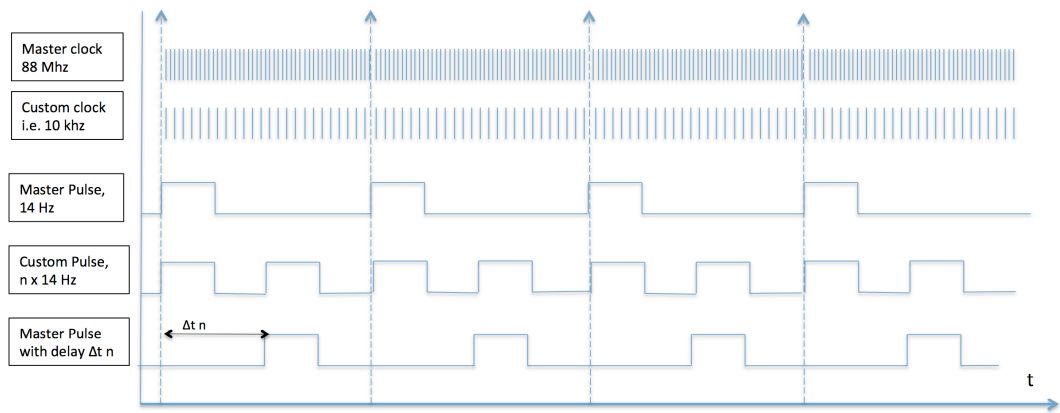

**Figure 2.** Set of synchronous strobe signals available at the timing receiver interface.

## 3. Instrument control and readout concept

### 3.1. Modular concept

The control and readout model that has been envisaged for ESS instruments is modular, where an instrument is made up from a number of independent subsystems ("modules") that do not interact with each other, but only centrally through the EPICS and DMSC interfaces. Each module simply takes care of it's own responsibilities, collecting data (which can be neutron or meta data) or controlling some physical configuration of the instrument (motion control, chopper speed and phase, magnets, etc.) or often both. These systems present the data in the natural form for the module – data is time-stamped, but the significance of the timing cannot be realised until the data is combined at the DMSC. Similarly, detector data is formatted in the natural units of the detector (channel number etc.) rather than physical units (e.g. position) so that the local readout systems do not have to change when an instrument is reconfigured.

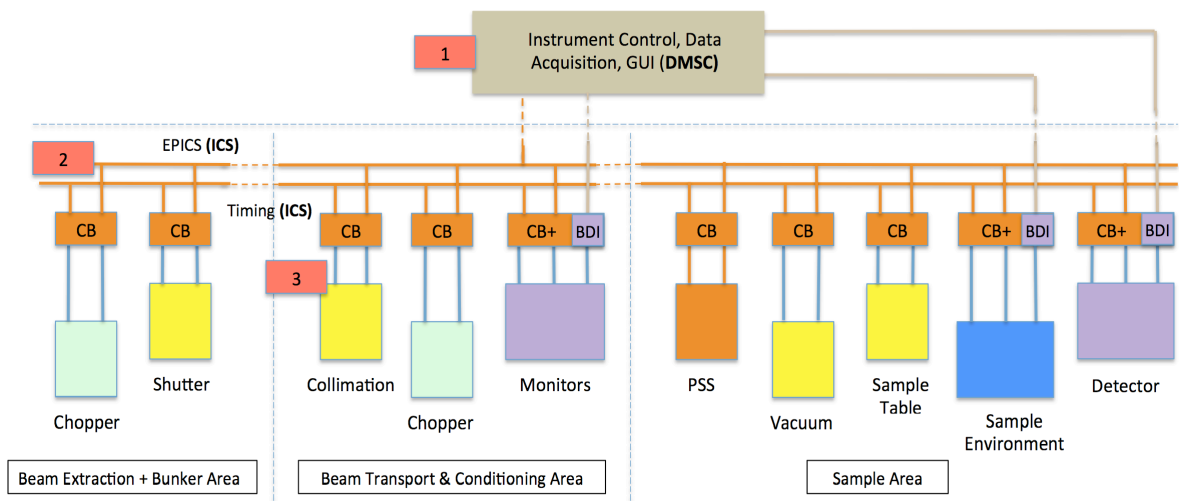

**Figure 3.** Topology of the modular instrument control concept for a 160m long Neutron Scattering Instrument at ESS. (CB = Control Box, BDI = Bulk Data Interface).

*3.2. Topology (figure 3)*
Backbone of the ESS instrument control is the EPICS and Timing network of the ICS with the control boxes as interface points (orange). Physically there will be one EPICS subnet per instrument with connectivity to the whole facility. All instruments subsystems (different colours according to functionality) link to these boxes and could be clustered in electrically isolated groups following the geographical layout of the instrument in the different instrument halls. The user interface of the DMSC (brown) is connecting to the EPICS layer via a data gateway. Modules with larger volumes of data like detector electronics or fast sample environment require a dedicated link to a DMSC aggregator node called a bulk data interface (BDI, violet, figure 3). The red boxes #1 to 3 indicate engineering user interfaces on the different layers for diagnostics purposes.

*3.3. Design principles*
Every control process that can be done (or has to be done) locally in one of the modules will be done locally, every signal that needs to be related to data from other modules has to be time stamped and sent to the DMSC user interface. That gives clear functionalities and clear interfaces for single modules, easy to specify and to bring in as in-kind contributions or from external suppliers. All modules are linked together by the instrument control/data acquisition software of the DMSC (via the EPICS layer); it's a crucial part of the instrument and will be tailored to the needs of each instrument. It gives the flexibility necessary to adapt the same hardware to each of the ESS instruments and to future instrument upgrades: the hardware is done once and prepared for all future (hi-level) functionalities.

However care must be taken about the latency in collecting the data from all modules together to allow the formation of a full 'frame' of neutron data being presented to the user for monitoring purposes. We will address this by introducing a maximum tolerated 'latency budget' as a design requirement for the whole instrument control system.

**4. Module interfaces and functionalities**
Each control module basically has two or three interface types to the upper device layers: Timing, command interface and – if necessary- an interface for larger volumes of data (figure 4).

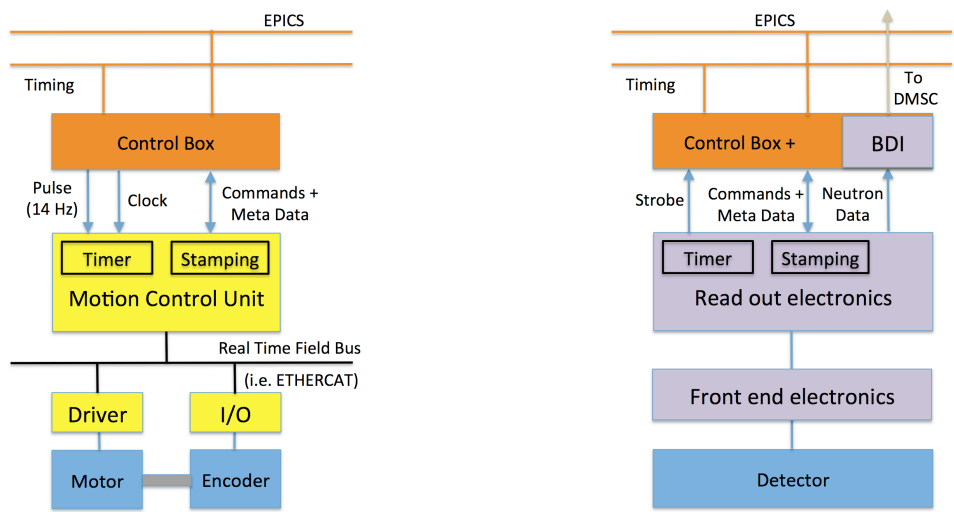

**Figure 4.** Interfaces and time stamping functionality for a motion control module (left) and a detector electronics module (right).

*4.1. Command and meta data interface*
The command interface connects through the control box to the EPICS network and is able to transfer smaller volumes of instruments metadata through this channel as well. This data is bridged via gateway to the DMSC user interface 'off instrument'. Currently the data rate is limited to about 1kB/s but improvements of data rates in later versions of EPICS (version 4) will be considered. Physically this command/meta data connection is mostly a serial interface like RS232 or Ethernet with latencies between 20 and 100 ms.

*4.2. Time stamping*
Following the modular control concept described in section 3 it is mandatory to timestamp all instruments data. This could easily be done by the timing receiver in the control box but is limited in precision by the latency of the connection between the modular control unit and the control box. For higher precision it is necessary to transfer the absolute time information from the control box to the control electronics of the module (or transfer the whole timing receiver). In figure 4 we describe two ideas of synchronising internal timing clocks in control modules with the absolute time of the timing receiver. Either a pulse from the control box is synchronising a timer in the control unit (motion control, left) or a strobe is sent from the control module to the control box and time stamped there (detector electronics, right). In both cases the absolute time of the event is sent to the module via the command interface where it can be related to the internal time of the module. Within the control module the absolute time information can be transferred further by means of real time field busses (e.g. ETHERCAT).

*4.3. Detector readout and bulk data interface (BDI)*
The readout electronics for detectors will be based on FPGAs that will perform data collection and transmission, and necessary (i.e. unavoidable) data reduction or matching. Although ESS has a high flux, the data rates are small compared to particle physics or radio astronomy and we are not too constrained by bandwidth concerns. We anticipate operating the BDI links at 10Gb/sec, but 40Gb/sec links are already readily available and affordable. To generalise, we expect to include basic background suppression in firmware, and cross data flagging to ensure all relevant data is output, but processing such as cluster centroid finding or time based data matching would be performed in software.

**5. Advanced use cases in neutron instrumentation**

*5.1. Taking neutron data on the fly*
Extending the event mode data acquisition from neutron data to motor positions is addressing the requirements for a 'scanning on the fly' in the most flexible way. Neutron data can be related to freely configurable motor position ranges; a binning that might be adapted after the experiment according to analytical or statistical needs. Even higher grades of automation are possible with this setup where the speed of an instrument movement is determined on the fly by the count rate and the desired statistics. At the same time the sampling rate of the motor position data is determined by the wanted precision in space and the given speed. Introducing synchronised movements with non-linear relations between axes even makes something like fully automated constant energy scans with neutron data on the fly feasible.

*5.2. Stroboscopic mode in sample environment*
Once the time stamping procedure is implemented in the instruments subsystems no further hardware development is necessary to introduce new time relation between neutron data and processes running on the instrument (e.g. in sample environment). Software development can be done 'off instrument' and it is easier to experiment with alternative processing schemes in parallel. At the same time processes can be synchronised to the 14Hz proton pulse if required by the scope of the experiment. Stroboscopic experiments are easy to set up as the properties of the process and the neutron data depend on exact the same time structure and therefor could be easily related.

*5.3. Flexible chopper veto*
Time stamping the TDC signal of all choppers allows for a flexible handling of the veto condition for data acquisition. Per default all experimental data is acquired and grouped/sorted in different quality classes according the value of the corresponding TDC signal. The user can decide after the experiment which 'quality' of data he includes in his experimental analysis. This might save usable data that otherwise would be lost due to strict veto conditions.

*5.4. Diagnostics of beam transport and chopper systems*
Advanced diagnostics is a key requirement of ESS instrumentation. We envisage placing a neutron beam monitor behind each chopper group to monitor the neutron flux. By time stamping the monitor events the level and the time structure of the neutron flux at the different positions of the beam line can be related to the time structure of the corresponding choppers TDC signals. Results can be presented on an 'oscilloscope style' diagnostics screen or on a time-distant diagram like figure 5. At the same time the properties of the beam transport system can be analysed and monitored by comparing the neutron flux at the corresponding opening times of the choppers (as indicated in figure 5 for the brown 2.5 - 3.97Å range).

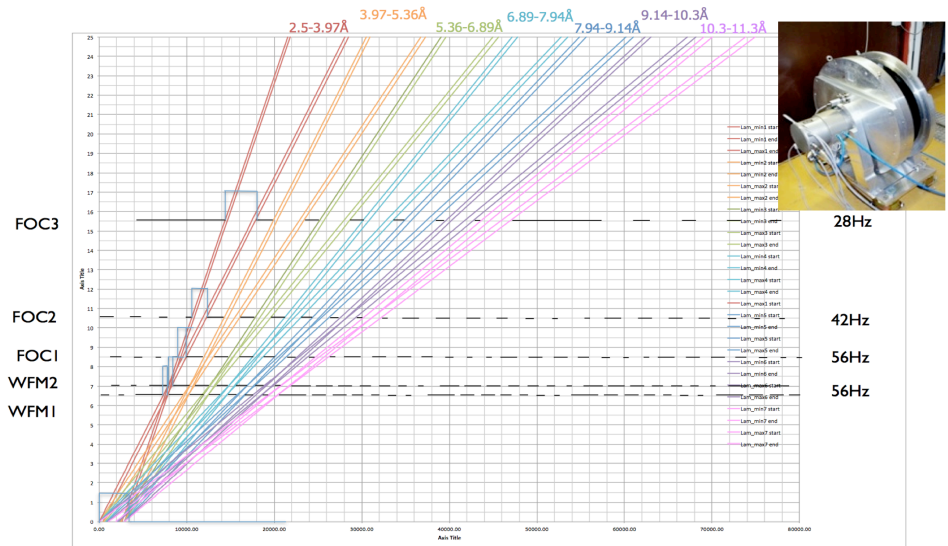

**Figure 5.** Time-distance (x-y) diagram of a wave frame multiplication chopper unit.

## 6. Conclusion
With the proposed modular instrument control concept we address the demanding challenges for ESS instrumentation. Instrument subsystems are simpler - easier to specify and easier to maintain. With clear defined functionalities and interfaces they are easy to bring in as in-kind contributions or from external suppliers. The proposed concept fits exactly in the topology of the ESS wide control infrastructure thus profiting from standardisation and maximising synergies within the whole ESS project.

In some cases the concept of an extended event mode data acquisition represents additional hardware requirements e.g. for time stamping or data transfer. Data rate and latency requirements have to be addresses carefully in the design phase. But once the hardware is standardised and in place the subsystems are prepared for the majority of the future functional and operational upgrades. The functionality of the whole system depends solely on the DMSC user interface software that might start with simple standard applications, introducing gradually more demanding functionalities in future upgrades.

Already now the concept is supporting advanced use cases for experiments and diagnostics like on-the-fly scans, stroboscopic data acquisition or advanced neutron beam diagnostics. A lot more will follow in the future once the ESS has started commissioning and user operation.

**Acknowledgments**
We would like to thank our colleagues and collaborators in and outside ESS for the lively and fruitful discussions and the contribution of ideas and possible solutions with a special mention of ISIS, PSI, JCNS and CERN.